\documentclass[amsmath,amssymb,draft
aps,prl,showpacs,twocolumn,superscriptaddress]
{revtex4-1}
\usepackage{graphicx}
\usepackage{dcolumn}
\usepackage{bm}
\usepackage{color}
\definecolor{mycolor1}{rgb}{0,0,1}

\begin{document}

\preprint{APS/123-QED}

\title{Magnetic-Field-Induced Polar Phase in Chiral Magnet CsCuCl$_3$}

\author{Atsushi Miyake}
\thanks{miyake@issp.u-tokyo.ac.jp}
\affiliation{Institute for Solid State Physics, The University of Tokyo, Kashiwa, Chiba 277-8581, Japan}
\author{Jumpei Shibuya}
\affiliation{Institute for Solid State Physics, The University of Tokyo, Kashiwa, Chiba 277-8581, Japan}

\author{Mitsuru Akaki}
\affiliation{Institute for Solid State Physics, The University of Tokyo, Kashiwa, Chiba 277-8581, Japan}
\affiliation{Center for Advanced High Magnetic Field Science, Graduate School of Science,
Osaka University, Toyonaka, Osaka 560-0043, Japan}

\author{Hidekazu Tanaka}
\affiliation{Department of Physics, Tokyo Institute of Technology, Meguro-ku, Tokyo 152-8551, Japan}


\author{Masashi Tokunaga}
\affiliation{Institute for Solid State Physics, The University of Tokyo, Kashiwa, Chiba 277-8581, Japan}


\date{\today}

\begin{abstract}
Magnetoelectric effects in the chiral magnet CsCuCl$_3$ have been investigated through the magnetization and electric polarization measurements in pulsed high magnetic fields.
If the magnetic field is applied normal to the spin chiral $c$-axis, a plateau and a jump in magnetization are observed.
Between the plateau and the jump in magnetization, a small but significant electric polarization of about 0.25~$\mu$C/m$^2$ along the $a$-axis is observed with the field applied almost perpendicular to the $ac$-plane.
In addition, the paramagnetoelectric effect expected from the point group symmetry is confirmed.
The emergence of the electric polarization is explained by the cooperation of the local electric polarization on the chiral solitonic spin arrangement and the paramagnetoelectric effect.
Hence, we interpret this novel multiferroic phase in CsCuCl$_3$ as a polar solitonic phase.

\end{abstract}

\maketitle



\begin{figure}[t]
   \begin{center}
      \includegraphics[width=86mm]{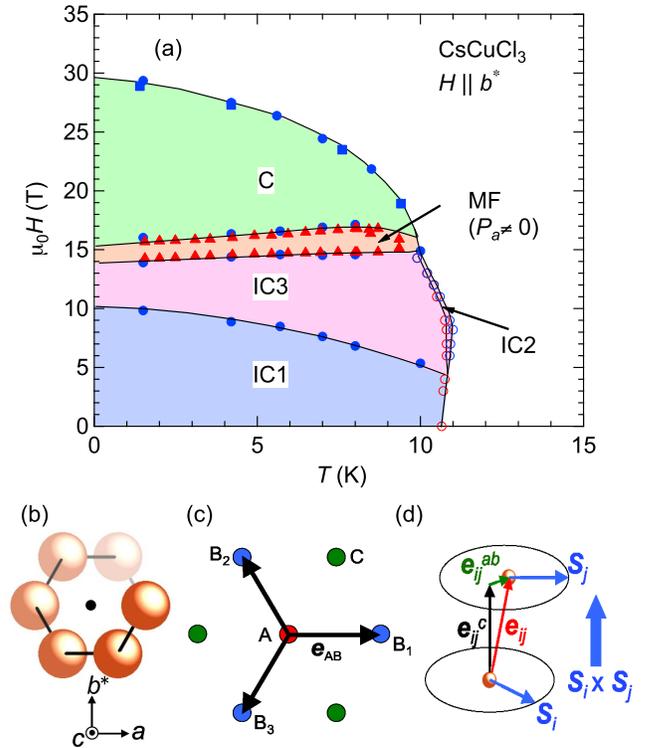}
   \end{center}
   \caption{(color online) (a) Magnetoelectric phase diagram of CsCuCl$_3$ in the $H_{b^{\ast}}$-$T$ plane.
Solid circles (triangles) denote the transition fields determined by the $M$-$H$ ($P_a$-$H$) curves.
Solid squares and open circles are taken from \cite{Shibuya2014, Stusser2002}.
See the text for details of each phase.
 (b) The Cu-chain viewed from the $c$-axis.
 The solid lines connecting the adjacent Cu-ions correspond to the $\vec{e}_{ij}$.
 The small circle indicates the $c$-axis of unit cell.
 (c) Triangular lattice consists of three-sublattices, sites A, B, and C.
 Arrows indicates $\vec{e}_{AB_i}$, where $i$~=~1, 2, and 3. 
 (d) Schematic of the nearest interlayer two Cu-ions.
 Arrows denote $\vec{e}_{ij} (= \vec{e}_{ij}^{ab} + \vec{e}_{ij}^{c}$) and $\vec{S}_{i(j)}$. 
The thick arrow indicates the direction of $\vec{S}_i\times \vec{S}_j$.
 }
   \label{fig:HTPD}
\end{figure}

Various triangular lattice antiferromagnets (TAFM) have long been studied with the focus on their geometric frustration from the antiferromagnetic coupling of the neighboring spins on a triangular lattice \cite{Achiwa1969, Collins1997, Tanaka2003}. 
Among the TAFM systems, a hexagonal CsCuCl$_3$ has received sustained interest because of its distinctive chiral crystal structure and its resultant magnetism under applied magnetic fields. 
The $S$ = 1/2 spins of the Cu$^{2+}$ ions form ferromagnetically coupled chains along the $c$-axis with $J_0$~=~28~K \cite{Hyodo1981}. 
The chains are antiferromagnetically coupled and form a triangular lattice in the $ab$-plane, leading to a 120$^{\circ}$ spin structure below $T_{\rm N}$ = 10.7~K at zero field \cite{Adachi1980}: the coupling strength is estimated to $J_1$~=~-4.9~K \cite{Tanaka1992}.
The Curie-Weiss temperature obtained by the high-temperature magnetic susceptibility is comparable to $T_{\rm N}$ for the magnetic fields along and perpendicular to the triangular lattice \cite{Achiwa1969, Tezuke1982}.   
One of the most striking characteristics in CsCuCl$_3$ is that the Cu-chains are displaced helically in the triangular plane because of the cooperative Jahn-Teller distortion below 423~K \cite{Hirotsu1977}: the Cu-chains form helices along the $c$-axis with a six-ion periodicity [see Fig.~\ref{fig:HTPD}(b)] \cite{Schlueter1966}.  
Depending on the chirality, CsCuCl$_3$ has space group of $P6_122$ or $P6_522$ (i.e., point group 622) \cite{Kousaka2009}.  
The absence of inversion symmetry makes the Dzyaloshinskii-Moriya (DM) interaction active: the strength is estimated to $D$~=~5~K \cite{Adachi1980}, which is comparable with $J_1$.
The ferromagnetic (FM) interaction tends to align spin moments, whereas the DM interaction favors perpendicular arrangements of adjacent spins.
The resultant spin configuration for CsCuCl$_3$ is helical arising from the competition of these interactions with a period of 11.8$c$ \cite{Adachi1980}.

\begin{figure}[t]
   \begin{center}
      \includegraphics[width=86mm]{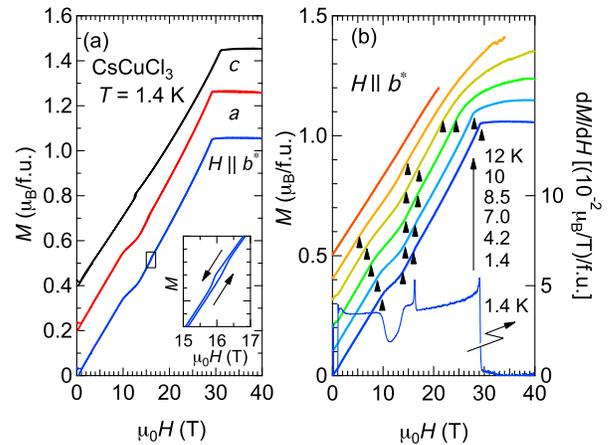}
   \end{center}
   \caption{(color online)(a) Magnetic fields ($H$) dependence of magnetization ($M$) for $H$ parallel to the $a$, $b^{\ast}$, and $c$-axes at 1.4~K.
Data are offset by 0.2~$\mu_{\rm B}$/f.u. for clarify.
The inset shows an enlarged view of the $M(H)$-curves for $H~||~b^{\ast}$, corresponding to the rectangle in the main panel.
(b) $M(H)$-curves at various temperatures (left axis) and d$M$/d$H$-curve at 1.4~K (right axis) with increasing fields for $H~||~b^{\ast}$.
The arrow heads indicate the phase transition fields, which are determined by the inflection points or peaks of d$M$/d$H$-curves.
The $M(H)$-curves are offset by 0.1~$\mu_{\rm B}$/f.u. for clarify.
}
   \label{fig:MH}
\end{figure}

The singularities of magnetization ($M$) in CsCuCl$_3$ were first observed by measurements in pulsed-magnetic fields \cite{Nojiri1988}; these are reexamined in this study (Fig.~\ref{fig:MH}).  
With the magnetic field ($H$) applied along the $c$-axis, $M$ shows a jump at 12.5~T and is saturated above 31~T.
This jump in $M$ for $H~||~c$-axis was theoretically explained by taking into account the quantum fluctuation as a transition from an umbrella structure to a coplanar one at 12.5~T \cite{Nikuni1993}.
In contrast, the effect of transverse field is more complicated. 
With an applied $H$ normal to the $c$-axis at 1.1~K, $M$ shows a linear $H$-dependence up to 10.5~T, then a plateau-like behavior between 10.5~T and 14.5~T, and finally a monotonic increase up to the spin saturation fields \cite{Nojiri1988}. 
To date, three incommensurate phases and a commensurate phase have been determined below the spin saturation fields through neutron scattering and thermodynamic measurements \cite{Stusser2002}.
The magnetic structure is expressed by an incommensurate magnetic propagation vector (1/3, 1/3, $\delta$), where $\delta$ is 0.085 at zero field \cite{Adachi1980}.
The $H$-dependence of $\delta$ has been well studied \cite{Mino1994, Nojiri1998, Stusser2002, Stusser2004}.
With increasing $H$, $\delta$ decreases to $\sim$0.05 at the beginning of the $M$-plateau \cite{Mino1994}.
This region is referred to as the IC1 phase \cite{Schotte1998, Stusser2002}.
In the plateau region, $\delta$ is locked to this value independent of temperature \cite{Mino1994, Stusser2002}.  
On further increasing $H$, $\delta$ decreases to 0 \cite{Nojiri1998, Stusser2002, Stusser2004}, entering in a commensurate (C) phase.
The spin spirals along the $c$-axis disappear in C phase. 
The field region separating the IC1 and C phases is called the IC3 phase \cite{Stusser2002}.
Just below $T_{\rm N}$ and an applied $H$, another incommensurate phase (IC2) with non-120$^{\circ}$ spin structure was discovered \cite{Schotte1998}.
Although there are several high-field experiments \cite{Nojiri1988, Mino1994, Nojiri1998, Stusser2002, Stusser2004} and theoretical studies \cite{Jacobs1993, Nikuni1998} on the magnetic transitions in fields normal to the $c$-axis, all aspects of these fascinating properties have not yet been revealed. 

In addition to the effect of quantum fluctuation in CsCuCl$_3$, we are interested in the aspect of helical spin-arrangements. 
Chiral magnets have recently attracted much attention because of novel electronic properties arising from their novel spin textures.
Interestingly, they can be finely controlled by applying a magnetic field.
For example, a metallic chiral magnet CrNb$_3$S$_6$ shows a fascinating chiral soliton lattice \cite{Togawa2012}.
CrNb$_3$S$_6$ has space group $P6_322$, which has the same point group as CsCuCl$_3$, and the chirality of the Cr-chains allows a DM interaction with the $D$-vector pointing to the chain.
Similar to CsCuCl$_3$, the interlayer Cr-ions are coupled via the FM interaction.
If $H$ is applied normal to the helical axis, $H$ tunes very precisely the periodicity of the soliton lattices, $L$ up to an $H$ aligning ferromagnetically the entire spins \cite{Togawa2012}.
The interlayer magnetoresistance along the helical axis is also well explained by the $H$-dependence of the soliton density, specifically, proportional to $L^{-1}$ \cite{Togawa2013}.
CsCuCl$_3$ with a similar easy-plane magnetic structure is also expected to form a chiral soliton lattice in the transverse magnetic fields.
Moreover, the incommensurate index $\delta$ of CsCuCl$_3$ may be regarded as the soliton density.
The $\delta(H)$-curve above the plateau \cite{Nojiri1998, Stusser2002, Stusser2004} is also reminiscent of the $H$-dependence of the soliton density, as observed in CrNb$_3$S$_6$ \cite{Togawa2012}.

In contrast to metallic systems, the insulating CsCuCl$_3$ may show novel magnetoelectric properties as indicated by the following expectations.
The magnetoelectric response corresponding to the second-order term in the magnetic field is non-zero for point group 622 of CsCuCl$_3$ \cite{Borovik2013}, e.g., the electric polarization ($P$) along the $a$-axis $P_a$ is proportional to $H_{b^{\ast}}H_c$, where $b^{\ast}$ is defined as the normal to the $ac$-plane.
This paramagnetoelectric effect emerges without spin ordering.
In addition, spin order reduces symmetry and may promote ferroelectricity, caused by the spin texture. 
Previously, we observed a small $P$ perpendicular to the $H$-direction along the triangular plane just above the magnetization plateau \cite{Shibuya2014}.  
In addition, the transition fields and $P$ strongly depend on the $H$ angle.
Moreover, the sign of $P$ changes with the $H$-direction crossing the $b^{\ast}$-axis, and $P$ vanishes when $H$ is precisely applied along the $b^{\ast}$-axis.
To reveal the emergence of the electric polarization, we investigate more details of multiferroic properties of CsCuCl$_3$ for the transverse field effect for which the magnetic fields are almost normal to the chiral $c$-axis.

\begin{figure}[h]
   \begin{center}
      \includegraphics[width=86mm]{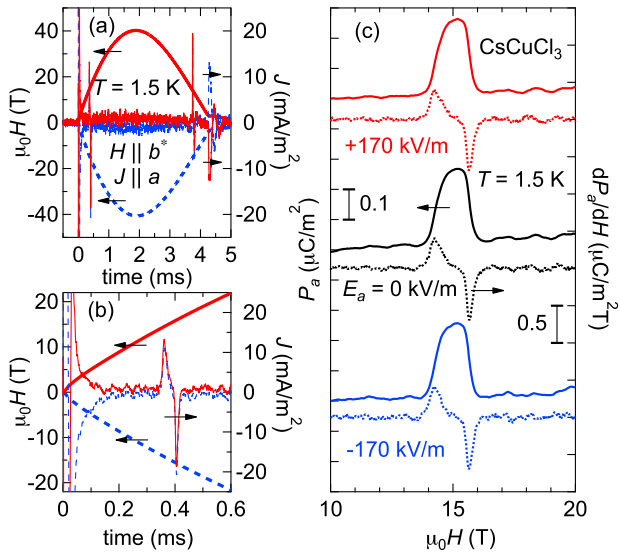}
   \end{center}
   \caption{(color online) (a) Magnetic-field polarity dependence of the polarization currents normalized by the measuring area, $J$, of $P_a$ of CsCuCl$_3$ at 1.5~K as a function of time with the positive (solid line) and negative (dashed line) $H$ profiles.
The thick and thin lines are $H$ (left axis) and $J$ (right axis), respectively.
   (b) Expanded view of (a).
   (c) $H$ dependence of $P_a$ of CsCuCl$_3$ with poling electric fields $E_a$ = 0 and $\pm$170~kV/m applied along the $a$-axis.
   For clarify, the data with increasing magnetic fields are only shown.
   The solid and dotted lines are $P_a$ (left axis) and $H$ derivative of $P$ (right axis), respectively. 
      }
   \label{PH_B_E}
\end{figure}

Single crystals of CsCuCl$_3$ are prepared by dissolving stoichiometric amounts of CsCl and CuCl$_2\cdot$H$_2$O in water, and the solution evaporates slowly.
The crystal axes are determined by the Laue diffraction method.
The accuracy of the crystallographic orientation is estimated to within about 1$^\circ$ by measuring rocking curves.
Pulsed magnetic fields are applied up to 56~T, using the non-destructive pulsed-magnets with typical durations of $\sim$ 4~ms or $\sim$ 8~ms installed at the International MegaGauss Science Laboratory of Institute for Solid State Physics at The University of Tokyo.
The profiles of the pulsed magnetic fields for the former one are shown in Fig.~\ref{PH_B_E}(a).
Magnetization was measured by the conventional induction method, using coaxial pick-up coils.
Changes in $P$ as a function of $H$ were estimated by numerically integrating the polarization current \cite{Mitamura2007}.

Figure~\ref{fig:MH}(a) represents $M$ as a function of $H$ applied along the $a$, $b^{\ast}$, and $c$-axes at $T$=1.4~K.
A jump at 12.6~T for $H~||~c$ and a plateau between 9 and 14~T for $H~||~a$ and $b^{\ast}$ are observed, which are in good agreement with the previous report \cite{Nojiri1988}.
The $M(H)$-curves are quite similar for in-plane applied fields, specifically along the $a$- and $b^{\ast}$-axes. 
In addition, for these $H$-directions a jump in $M$ at around 16~T is observed, which is clearly seen as a sharp peak in d$M$/d$H$ [Fig.~\ref{fig:MH} (b)].
A small $H$-hysteresis is observed in the inset of Fig.~\ref{fig:MH} (b), indicating a first-order phase transition.
At this field, the spin modulation wavenumber $\delta$ becomes zero, and therefore the spin spiral disappears \cite{Nojiri1998, Stusser2002}. 
Note that the upper bound field of the plateau, where $\delta$ starts decreasing, corresponds to the phase boundary between the paraelectric and ferroelectric phases, to be discussed later.  
Figure~\ref{fig:MH}(b) represents the $M(H)$-curves at various $T$s.
As indicated by the arrow heads, the plateau region expands as a function of temperature, and the jump is shifted to higher $H$.
These results are summarized as a $H_{b^{\ast}}$-$T$ phase diagram together with the $P(H)$ results in Fig.~\ref{fig:HTPD}(a).  
Hereafter, we shall focus on the magnetoelectric effect.

\begin{figure}[h]
   \begin{center}
      \includegraphics[width=86mm]{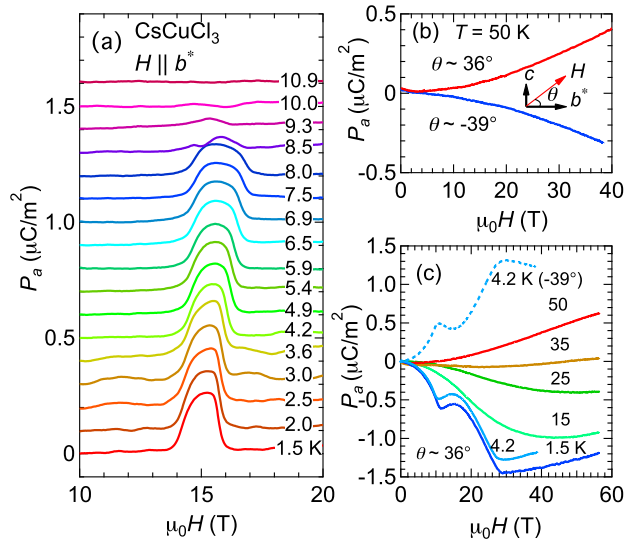}
   \end{center}
   \caption{(color online) (a) $H$-dependence of $P_a$ with $H~||~b^{\ast}$ at various temperatures for $H$-increasing processes.
The data are offset by 0.1~$\mu$C/m$^2$ for clarify.
(b) $P_{a}(H)$-curves for the $H$-angle $\theta\sim$~-39$^{\circ}$ and $\sim$~36$^{\circ}$ in the paramagnetic region $T$~=~50~K.
$\theta$ is the tilting angle from the $b^{\ast}$- to the $c$-axis.
The $H$-increasing processes are drawn for clarify.
(c) $P_{a}(H)$-curves with increasing $H$ for $\theta$ = 36$^{\circ}$ at various temperatures.
A curve for $T$=~4.2~K and $\theta$ =-39$^{\circ}$ is also plotted (broken line).
}
   \label{fig:PH}
\end{figure}

From Fig.~\ref{PH_B_E}, $P_a$ is induced between about 14 and 16~T for $H$ applied along the $b^{\ast}$.
To clarify the ferroelectric properties, we investigated the magnetic and electric fields effect on $P_a$.
Figure~\ref{PH_B_E}(a) shows typical time-profiles of the bipolar pulsed magnetic fields and polarization current normalized by the measuring area of approximately 6~mm$^2$, $J$.
Some peaks in $J$ are clearly visible.
In Fig.~\ref{PH_B_E}(b), sharp positive and negative peaks are observed at around 0.35~ms (13.9~T) and 0.4~ms (16~T), and do not change sign on changing the $H$-polarity.
This is clear evidence that the observed peaks in $J$ at these $H$s arise from the $H$-induced $P$.
In contrast, the extrinsic signal caused by inductive noise from rapid evolution of $H$ is also seen at the beginning of a pulsed magnetic field, which changes its sign with the $H$-polarity.
This is removed by averaging $J$ for the positive and negative $H$.
As shown in Fig.~\ref{PH_B_E}(c), the poling electric fields along the $a$-axis do not affect $P$, that is the magnitude and sign of $P_a$.
These results indicate that the small induced $P$$\sim$0.25~$\mu$C/m$^2$ observed in CsCuCl$_3$ is intrinsic and is not diminished by cancellations among the ferroelectric domains with degenerate spin configurations.
Because $P_a$ vanishes for $H$ applied precisely along $b^{\ast}$ \cite{Shibuya2014}, the observed finite $P_a$ indicates a misalignment of $H$ from the $b^{\ast}$-axis.
Figure~\ref{fig:PH}(a) represents the temperature variation of the $P_a(H_{b^{\ast}}$)-curves.
At 1.5~K, $P_a$ is induced between 13.9 and 16~T.
On warming, the amplitude of $P$ decreases with widening of the $H$-region for finite $P$.
In approaching $T_{\rm N}$, this tendency is more pronounced.
Within our experimental accuracy, $P$ disappears above 9.3~K.

Figure~\ref{fig:HTPD}(a) is the magnetoelectric phase diagram for the $H_b^{\ast}$-$T$ plane.
In addition to the previous phase diagram \cite{Stusser2002}, we unambiguously put a multiferroic (MF) phase in the IC3 from our $M$ and $P$ measurements. 
The MF phase is located between the upper bound $H$ of the $M$-plateau and $H$ entering the C phase.
The change in the IC vector $\delta(H)$ from constant to decreasing just above the $M$-plateau may play an important role in inducing the macroscopic $P$ (discussed below).

To reveal the further magnetoelectric properties of CsCuCl$_3$, we investigated the paramagnetoelectric effect.
In accordance with group theory analysis, the point group of CsCuCl$_3$ (622) dictates non-zero components for $P_a$ ($P_b^{\ast}$) for $H_b^{\ast}H_c$ ($H_aH_c$) \cite{Borovik2013}.  
To apply the magnetic fields along both the $b^{\ast}$- and $c$-axis directions, the sample used for $P_a(H_{b^\ast})$ measurements was tilted $\theta$ from the $b^{\ast}$- to $c$-axis.
As clearly seen in Fig.~\ref{fig:PH}(b), $P_a$ at 50~K well above $T_{\rm N}$ varies as a function of $H$ and angle $\theta$.
$P_a$ shows nearly symmetric $H$-variations for $\theta$$\sim$$36^{\circ}$ and $\sim$$-39^{\circ}$, which is clear evidence of the paramagnetoelectric effect.
Crossing $\theta$~=~0 ($H$~$||$~$b^{\ast}$), the $H$-component with respect to the $c$-axis changes polarity, and thus reversing $P_a$.
The change in $P_a (H)$ is comparable to that in the MF phase [Fig.~\ref{fig:PH}(a)].
Figure~\ref{fig:PH}(c) gives $P_{a} (H)$-curves for $\theta\sim 36^{\circ}$ at various temperatures.
The $P_{a}(H)$ strongly varies with $T$.
At 1.5~K, a sharp kink and a broad peak anomaly appear around 10~T and 15.3~T, respectively.
At $H\sim$28~T, the $P(H)$-curve shows a kink and changes its curvature.
The anomalies disappear above $T_{\rm N}$. 
While the origin of the anomalies are unclear, it seems to correspond to the $H$-induced magnetic phase transition.
The anomaly for $\theta\sim$-39$^{\circ}$ also shows a sign change. 
Surprisingly, the change in the curvature still remains above $T_{\rm N}$. 
At 15~K, for example, $P_a$ shows a minimum around 40~T.  
This is not expected simply from the paramagnetoelectric effect because it is only pertinent to the crystalline structure symmetry.
It may reflect the singularity of quantum fluctuation and/or short range spin correlation in CsCuCl$_3$.
The magnetic contribution of optical birefringence was reported to show a broad maximum at about 32~K \cite{Hyodo1981}.
This behavior is reproduced by taking into account the effect of short-range spin correlations arising from interchain interactions.
From this study, $P_a(H)$ changes to monotonic $H$-variation above 35~K [Fig.~\ref{fig:PH}(c)], but we have no clear explanations at present.
Further experimental investigation is required that provides more precise $\theta$- and $T$-dependences and clearer theoretical interpretation.

Recently, a similar $M(H)$-curve to that of CsCuCl$_3$ was reported in another $S$~=~1/2 easy-plane TLAF, Ba$_3$CoSb$_2$O$_9$ \cite{Susuki2013}.
In contrast to CsCuCl$_3$, the interlayer coupling of Ba$_3$CoSb$_2$O$_9$ is antiferromagnetic.
With increasing in-plane field, a small jump in $M$ was observed above a 1/3-plateau of the saturated $M$.
The jump is well reproduced with a model that includes a small antiferromagnetic coupling between the layers \cite{Yamamoto2015}.
The anomaly in $M$ corresponds to a first-order transition between the different stacking coplanar spin configurations.
For CsCuCl$_3$, the neighboring interchain spins are canted because of the competition between the FM and DM interactions.
Here, let us discuss whether the electric polarization can arise from the spin order in the triangular layer based on the spin-current model: $P= \sum_{j}\vec{e}_{ij}\times(\vec{S}_i \times \vec{S}_j)$, where the two spins $\vec{S}_{i(j)}$ are connected with a unit vector $\vec{e}_{ij}$ \cite{Katsura2005}.
For CsCuCl$_3$, the magnetic Cu-ions lie on the triangular lattice.
If the spin order can be represented by fixed vectors on three sub-lattices, the local $P$ cancels out, i.e., 
$\sum_{j=1}^{3} \vec{e}_{{\rm AB}_i} \times (\vec{S}_{\rm A} \times \vec{S}_{{\rm B}_j})=(\vec{S}_{\rm A} \times \vec{S}_{\rm B})\times(\sum_{j=1}^3\vec{e}_{{\rm AB}_i})=0$ in Fig.~\ref{fig:HTPD}(c), regardless of the direction of the spin vectors.
In other words, we cannot distinguish the in-plane magnetic structure through the $P$ measurements.
Hereafter, we shall focus on the interlayer spin-arrangement.

Let us recall the characteristics of the crystal structure of CsCuCl$_3$ again.
A chiral Cu-chain viewed from the $c$-axis is drawn in Fig.~\ref{fig:HTPD}(b).
Cu-ions are displaced helically with six-ion periodicity.
The unit vector connecting the resultant between the adjacent Cu-ions, $\vec{e}_{ij}$, is not parallel to the $c$-axis.
As shown in Fig.~\ref{fig:HTPD}(d), the in-plane component $\vec{e}_{ij}^{ab}$ is finite and is perpendicular to the $\vec{S}_i\times\vec{S}_j$.
This configuration allows a local finite $P$ between the neighboring magnetic-ions.
This differs strikingly when compared with the proper-screw-type magnetic structure in a straight chain, where the $\vec{S}_{i}\times \vec{S}_j$ is parallel to $\vec{e}_{ij}$, and hence for the spin-current model, the macroscopic polarization vanishes.
Even in the helical crystal, the summation of the local $P$ in the unit cell vanishes as long as the spin rotates uniformly.
This is the reason why no macroscopic $P$ is observed in the IC1 phase.

With increasing $H$, higher harmonics appear in the spin modulation \cite{Stusser2002, Stusser2004}.
In this situation, the total $P$ in the crystallographic unit cell does not necessarily cancel out.
The period of spin modulation is determined by the competition between the FM and DM interactions together with quantum fluctuations up to the plateau region in the IC3 phase. 
If the period of the spin modulation is incommensurate with that of the lattice, the total $P$ over the entire chain will cancel out again. 
We assume this is the reason why we do not observe finite $P$ in the plateau region.

With further increasing $H$, spin modulation period rapidly increases \cite{Nojiri1998, Stusser2002}.
According to the classical theoretical calculation \cite{Jacobs1993}, the spin modulation becomes significantly anharmonic in this field region so that it can be regarded as a soliton lattice in the long period limit.
As mentioned above, the spin-current model predicts the emergence of local $P$ at the soliton.
Because the distance between adjacent solitons is large, we expect that the location of the soliton in the Cu-chain can be easily modified by the external perturbation.
If this perturbation aligns with the local $P$ by changing the position of the solitons along the chains, experimentally we will observe a macroscopic $P$.

Although we cannot uniquely determine the origin at moment, we can discuss the role of paramagnetoelectric effect on symmetry breaking.
As shown in Figs.~\ref{fig:PH}(b) and (c), the superposition of finite $H_c$ and $H_{b^{\ast}}$ determines a preferable direction for $P_a$ through the paramagnetoelectric effect, and hence, can align the local $P$. 
Our previous field-angle dependence of $J$ measurements also support this scenario.
The sign change in $J$ with the field direction crossing the $b^{\ast}$-axis is clearly observed \cite{Shibuya2014}.
Importantly, $J$ is not induced if $H$ is applied {\it exactly} along the $b^{\ast}$-axis.
This fact indicates that the local $P$ for the soliton is not forced to align with $H$ applied parallel to the $ab^{\ast}$-plane, in agreement with our interpretation, i.e., the paramagnetoelectric effect determines the $P$ direction.
The observation of $P_a$ for $H~||~b^{\ast}$ [Figs.~\ref{PH_B_E} and \ref{fig:PH}(a)] is due to the small misalignment of the $H$-direction from the $b^{\ast}$-axis.
The tiny induced $P$ is also consistent with the small $S$ and $e_{ij}^{ab}$, and the weak paramagnetoelectric effect.
Therefore, we interpret this novel MF phase in CsCuCl$_3$ as the polar solitonic phase.

In conclusion, we have studied the magnetization and electric polarization of a chiral triangular lattice antiferromagnet CsCuCl$_3$ in pulsed magnetic fields.
For transverse magnetic fields, a tiny macroscopic electric polarization is induced from just above the magnetization plateau to the commensurate transition field, if the period of the incommensurate spin modulation is sufficiently lengthened.
These multiferroic properties are reasonably explained by assuming the polar soliton lattice state in this chiral magnet.
Finding a magnetoelectric effect in other chiral insulating magnets with larger spins will be of interest.
   
\begin{acknowledgments}
The authors are grateful to S. Miyahara, Y. Kousaka, and J. Kishine for fruitful discussions.
Sample alignments were performed using X-ray laboratory of ISSP, The University of Tokyo with the help of T. Yajima.
This work was partially supported by the MEXT of Japan Grants-in-Aid for Scientific Research (15K05145, 25610087 and 26247058).
\end{acknowledgments}

\end{document}